# Socioeconophysics: Opinion Dynamics for number of transactions and price, a trader based model

Çağlar Tuncay
Department of Physics, Middle East Technical University
06531 Ankara, Turkey
caglart@metu.edu.tr

**Abstract**

Involving effects of media, opinion leader and other agents on the opinion of individuals of market society, a trader based model is developed and utilized to simulate price via supply and demand. Pronounced effects are considered with several weights and some personal differences between traders are taken into account. Resulting time series and probabilty distribution function involving a power law for price come out similar to the real ones.

## 1. Introduction

Members of market society are well known to act fundamentally according to their opinion [1] about to buy, to sell, and to wait, rather than real economic criteria. Opinions effect others and are effected by them and by other agents such as, media and opinion leaders, during their evolution through people, which are modeled as being cited on a square matrix. One of these approaches is known as the Sznajd model[2]. (Relevant issues may be found in [3]). It was amply used in attempts to predict votings. (See [4-11] for 'voting model', and see [11-15] for applications in some surface effects, such as catalytical reactions in material science. And, for economy and market related issues of opinion dynamics, see [15-17], and references therein.)

In the following section we apply opinion dynamics to market, and in the next one we display our simulation results. Last section is devoted to conclusion.

## 2. Opinion Dynamics for Market Society

The present society is composed of ten thousand people, each willing to gain money. Yet, this characteristics does not unify but splits them into three categories as the buyers (demanders, bulls), the sellers (suppliers, bears), and the waiters (wolfs). Each of them seeks for the opinion of the others, since the market is made anonymously; but each may react to the gathered (surrounding) opinion in different manners. For example, in many cases buying of others (demand) prompts selling (supply) of an individual, and vice versa. And from time to time herding mechanism works. Market customers are not like the referees of a beauty contest and atoms of a crystal. In the present approach, a random individuality is assigned to each member, as a basic difference with respect to the previous applications of opinion dynamics to different societies. Secondly, for each entry of the matrix, the four nearest neighbors are separated from the next nearest neighbors in the strength of bonds inbetween. Because, the ability of one member to convince any other one may change from person to person. On the other hand, market people do not always believe to what they hear. Moreover, they may even do the opposite of what they are adviced to do. And indeed, there exist many people within exchange who prefer to buy (strategically) while the others are selling (i.e., during recessions), and vice versa.

Individual opinions are initially let randomly to be either -1 (sell), or +1 (buy) or 0 (wait), i.e. each by 33.33% but which habitant has what opinion is selected randomly in the present approach. So, initial demand equals to the supply. As interaction tours take place, in a manner described in [2-4, and 16], the range of opinion effection extends. After few runs, a considerable number of the 100x100 matrix inhabitants become effected by a considerable number of co-matricians. Accordingly, excess occurs within the numbers of transactions demanded and supplied, which creates an impact on the price. For example, if for a given price level, the number of demanded transactions becomes greater than the supplied (booked) amount, price increases. Because, the present booked ones will be bought and the excess amout (equals to demand minus supply) have to be bought from the level which is one step higher in price, and vice versa. It is well known that, price of shares change in discreate amounts (steps) in real markets. Some portion of money composing the volume is spent to increase the number of transactions processed with a fixed price, and the other portion is spent to change the price. (See [18 and 19] and references therein.)

Negotiators looking for a compromise, usually follow advices of their broker or some commentator; for example, an experienced and skillful friend or relative. Such an opinion leader (market guru), is cited with fixed coordinates in the bulk, i.e., far from the borders of the society matrix. (See, references in [3].) The effect of the opinion leader is kept constant throughout the tours, which may be varied in other tours. Mass media, trying to influence opinions through advertising may also be handled in a similar manner, with its effect covering the whole inhabitants at any given time. It was found [20] that, "The larger the lattice is the smaller is the amount of advertising needed to conceive the whole market."

Each market morning, people come to market with three possible opinions in their mind; to buy, to sell, or to wait (watch). And accordingly, they process or wait. In the present model, we take 33.33% of the society as buyers (with opinion equals to +1), and 33.33% of the society as sellers (with opinion equals to –1), and the rest (33.33%) as waiters initially. As time goes on, their opinions may change during the interaction tours. And in coming tours, following the evolution process described in [2-4, and 16], some may become new buyers, some may become new sellers, and some may become new waiters.

## 3. Market Simulations

To sharpen the notation followed here: We call a collection of few runs (say, 4 in number) 'a day', and at the beginning of each new day we set the whole society to its initial values, i.e. the opinions to be either -1, or +1, or 0, as described in the previous section. The rest of the parameters are all kept as the same, except the leader or media effect, if any is being studied.

We figured out that, the leader or media effect become important in medium run, and any may not be regarded within our days comprising 4 tours. Figure 1 exemplifies an evolution of a leader (cited at the center) effect (with relative strength equals 0.8, white in colour in Fig.1. a) through the society. Within the first 5 tours (Fig.1. b), the leader effect spreads over her neighbors and simultaneously many other members become induced to have opinions ranging between 0.75 and 1, due to natural interactions. And at the end (after the next 5 tours) the pronounced effect becomes completely intracable (Fig.1. c), as can be compared with the natural distribution after 10 tours (Fig.1. d). According to the present result, one needs more than one agent to speculate on the market. And, short-cut is the mass media, where we obtain the price growing exponentially in tours (not shown), even with small values of relative strength of influence.

In a scientific society there may not exist any antithetical member, yet in exchanges there exist many, especially within dealers and medium-big portfolio owners. Moreover, anybody may react oppositely from time to time. (Also, the whole amount of sellings are bought (or, vice versa) during herding.) We count all these to be one part in five (20%) upmost, in our market simulations. Figure 2 displays a 500 day (and 4 runs per day) result of a free interaction of opinions, i.e., without any guru and media effects. There, we utilized the threshold values of +0.95 and –0.95 for bullish and bearish processes, respectively. (See, Eq. (4) and the relevant text in [21] for details.) Please notify that, the number of transactions per trader processed at any price level is irrelevant here.

The reader may regard that, the shape of our simulated price chart (Fig.2. a.) is very similar to real ones, where columns below represent excess in transactions. Probability distribution function for fluctuations in price is diplayed in Fig.2. b., where we have a power law. The extraordinary tail in Fig.2. b is due to setting the society to our standard initials, each 'morning', when very stiff rises and falls may take place.

## 4. Conclusion

We run many more simulations with different parameters, and obtained very similar results. It is clear that, empirical values of coordination and strength of bonds, as well as the percentage of antithetical people in the given society are needed for better utilization of the presented results.

**FIGURE CAPTIONS**

**Figure 1.** An evolution of a leader effect (with relative strength equals 0.8) through the society. a- At the beginning of a day. **b-** After 5 tours. **c-** After 10 tours. **d-** After 10 tours, when no leader exist, i.e., with the relative strength equals zero.

**Figure 2.** **a-** 500 day (with 4 runs per day) result for price (X, with arbitrary unit) of a free interaction of opinions, i.e., without any guru and media effects. There, we utilized the threshold values of +0.95 and –0.95 for bullish and bearish processes, respectively. Below is the excess in number of transactions. **b-** Power law in price fluctuations.

**FIGURES**

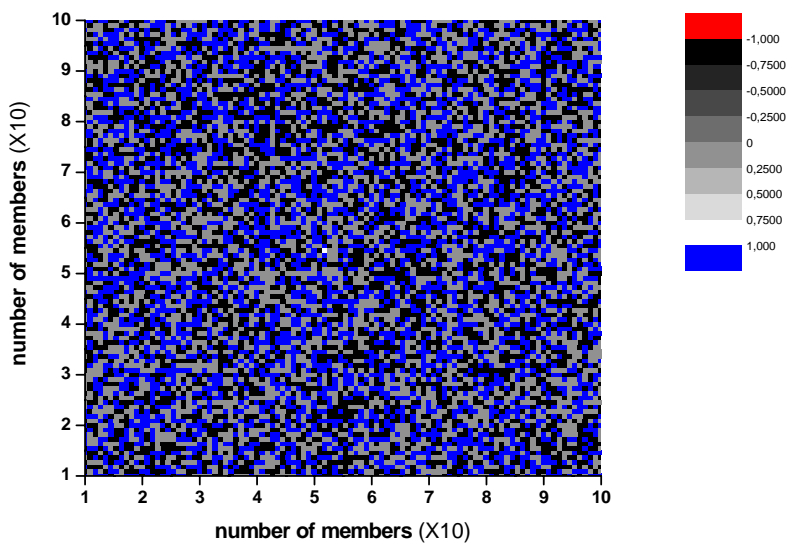

**Fig.1. a.**

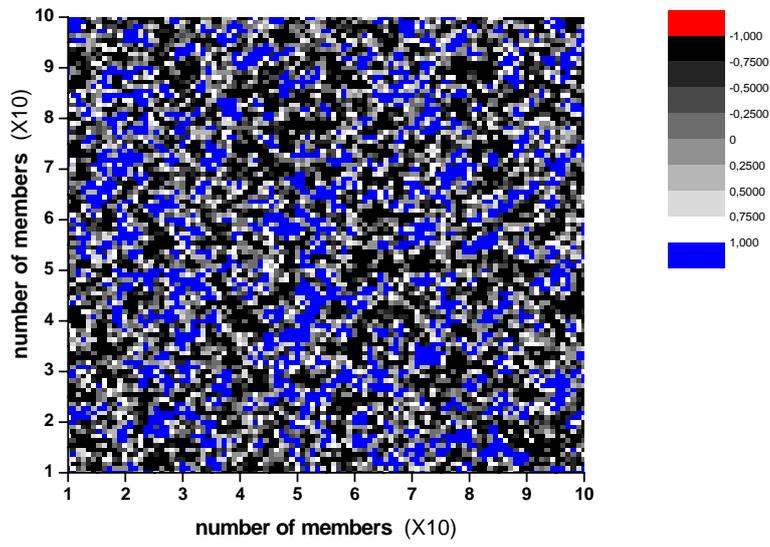

**Fig. 1. b.**

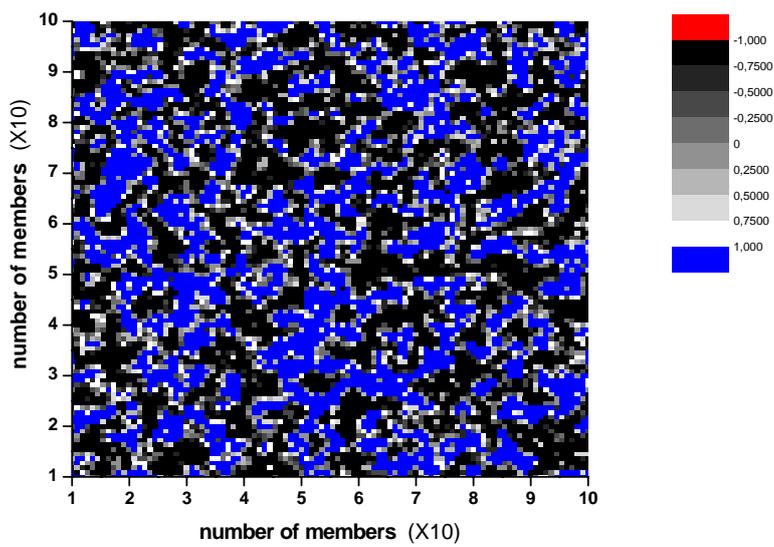

**Fig. 1. c.**

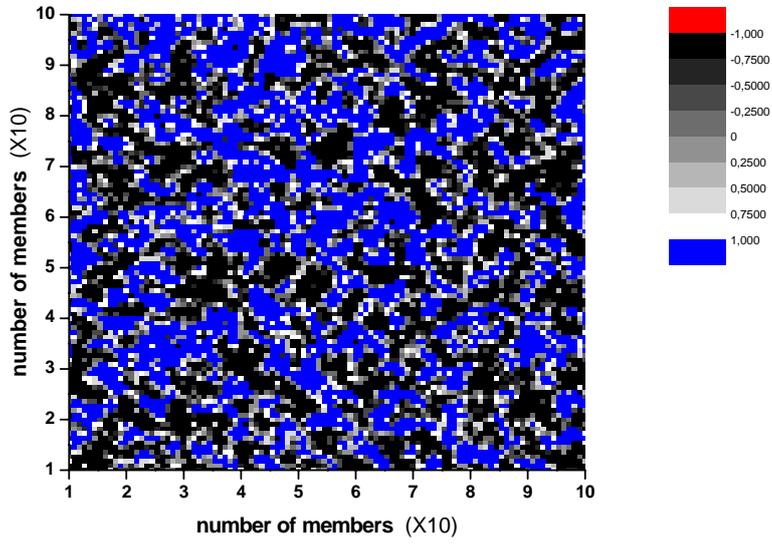

**Fig.1. d**

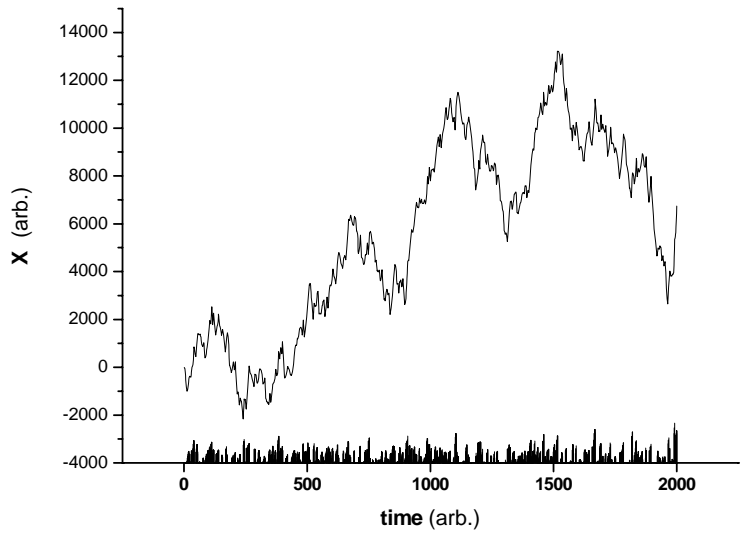

**Fig. 2. a**

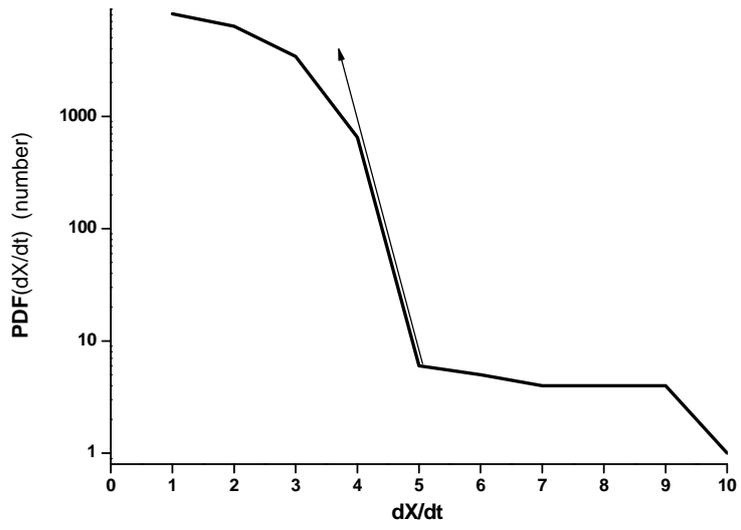

**Fig. 2. b**